\documentclass[traditabstract]{aa}

\usepackage{graphicx}
\usepackage{txfonts}
\usepackage{natbib}
\usepackage[colorlinks=true,citecolor=blue]{hyperref}
\usepackage{xspace}
\usepackage{array}
\usepackage[labelformat=simple]{caption,subcaption}
\usepackage{bm}
\usepackage{amsmath}
\usepackage{txfonts}
\usepackage{ulem}
\usepackage{multirow}
\usepackage{soul}

\bibpunct{(}{)}{;}{a}{}{,}

\newcommand{\lsd}{\hbox{$\lambda/D_{pup}$}\xspace}
\newcommand{\up}[1]{\textsuperscript{#1}}

\usepackage{color}

\defcitealias{Pourcelot2022}{Paper I}

\begin{document} 

\title{Low-order wavefront control using a Zernike sensor \\ through Lyot coronagraphs for exoplanet imaging:}
\subtitle{II. Concurrent operation with stroke minimization}

\author{
    R.~Pourcelot\inst{\ref{oca}}
    \and
    E.~H.~Por \inst{\ref{stsci}}
    \and
    M.~N'Diaye\inst{\ref{oca}} 
    \and
    H.~Benard \inst{\ref{tas}}
    \and
    G.~Brady \inst{\ref{stsci}}
    \and
    L.~Canas \inst{\ref{tas}}
    \and
    M.~Carbillet\inst{\ref{oca}}  
    \and \\
    K.~Dohlen \inst{\ref{lam}}
    \and
    I.~Laginja \inst{\ref{obspm}}
    \and
    J.~Lugten \inst{\ref{stsci}}
    \and
    J.~Noss \inst{\ref{stsci}}
    \and
    M.~D.~Perrin \inst{\ref{stsci}}
    \and
    P.~Petrone \inst{\ref{hf}}
    \and
    L.~Pueyo \inst{\ref{stsci}}
    \and \\
    S.~F.~Redmond \inst{\ref{Princeton}}
    \and
    A.~Sahoo \inst{\ref{stsci}}
    \and
    A.~Vigan \inst{\ref{lam}}
    \and
    S.~D.~Will \inst{\ref{gsfc}}
    \and
    R.~Soummer \inst{\ref{stsci}}
}

\institute{
    Université Côte d’Azur, Observatoire de la Côte d’Azur, CNRS, Laboratoire Lagrange, France \\
    \email{\href{mailto:raphael.pourcelot@oca.eu}{raphael.pourcelot@oca.eu}} \label{oca}
    \and
    Space Telescope Science Institute, 3700 San Martin Drive, Baltimore, MD 21218, USA \label{stsci}
    \and
    Thales Alenia Space, 5 Allée des Gabians - B.P. 99 - 06156 Cannes la Bocca Cedex – France \label{tas}|
    \and
    Aix Marseille Université, CNRS, CNES, LAM (Laboratoire d’Astrophysique de Marseille) UMR 7326, 13388 Marseille, France
    \label{lam}
    \and
    LESIA, Observatoire de Paris, Université PSL, Sorbonne Université, Université Paris Cité, CNRS, 5 place Jules Janssen, 92195 Meudon, France \label{obspm}
    \and
    Hexagon Federal, Chantilly, VA 20151, USA \label{hf}
    \and 
    Department of Mechanical and Aerospace Engineering, Princeton University, Princeton, NJ 08540, USA \label{Princeton}
    \and
    NASA Goddard Space Flight Center, Greenbelt, MD 20771, USA \label{gsfc}
}

\date{\today}

\abstract
    {Wavefront sensing and control (WFSC) will play a key role in improving the stability of future large segmented space telescopes while relaxing the thermo-mechanical constraints on the observatory structure. Coupled with a coronagraph to reject the light of an observed bright star, WFSC enables the generation and stabilisation of a dark hole (DH) in the star image to perform planet observations.}
    {While WFSC traditionally relies on a single wavefront sensor (WFS) input to measure wavefront errors, the next generation of instruments will require several WFSs to address aberrations with different sets of spatial and temporal frequency contents. The multiple measurements produced in such a way will then have to be combined and converted to commands for deformable mirrors (DMs) to modify the wavefront subsequently.}
    {We asynchronously operate a loop controlling the high-order modes digging a DH and a control loop that uses the rejected light by a Lyot coronagraph with a Zernike wavefront sensor to stabilize the low-order aberrations. Using the HiCAT testbed with a segmented telescope aperture, we implement concurrent operations and quantify the expected cross-talk between the two controllers. We then present experiments that alternate high-order and low-order control loops to identify and estimate their respective contributions.}
    {We show an efficient combination of the high-order and low-order control loops, keeping a DH contrast better than $5\times10^{-8}$ over a 30\,min experiment and stability improvement by a factor of 1.5. In particular, we show a contrast gain of 1.5 at separations close to the DH inner working angle, thanks to the low-order controller contribution.}
    {Concurrently digging a DH and using the light rejected by a Lyot coronagraph to stabilize the wavefront is a promising path towards exoplanet imaging and spectroscopy with future large space observatories.}


\titlerunning{Low-order wavefront control using a Zernike sensor through Lyot coronagraphs for exoplanet imaging II}

\maketitle

\section{Introduction}
High-contrast imaging and spectroscopy is one of the pathways envisioned by the Astro2020 Decadal Survey \citep{2020DecadalSurvey} to identify Earth-like worlds in other planetary systems and search for the biochemical signatures of life. To this aim, a telescope and its instruments need to fulfill at least the following three main requirements: sensitivity to observe objects with magnitude larger than 30, resolution to separate a planet from its host star at angular distances shorter than 50\,mas, and contrast to disentangle the planet photons from the star glow to observe exo-Earths around Sun-like stars with a contrast (i.e., flux ratio) around 10\up{-10} in visible light.  

A space telescope with a primary mirror larger than 4\,m \citep{Habex2019,LUVOIR2019} will have the resolution and sensitivity to enable the observation of a large sample of Earth-like planets \citep{Stark2019}. To save weight and room in a spacecraft and rocket fairing, a promising option for the telescope is to be folded thanks to the use of a segmented primary mirror, such as the recently launched James Webb Space Telescope \citep{Lightsey2012}. To overcome the flux ratio between the host star and the planet, one of the promising options is the use of a coronagraph. This instrument rejects the on-axis stellar light while retaining the planet light. In the case of geometrically unfriendly apertures of telescopes with pupil features such as primary mirror segmentation, secondary mirror central obstructions, and spider struts, the Lyot-style coronagraph \citep{Lyot1939} is one of the leading devices for starlight diffraction suppression with arbitrary apertures \citep[e.g.][]{Soummer2005, Soummer2011, N'Diaye2016a}. Including the combination of a focal-plane mask and a Lyot stop, this concept has already been implemented in several ground-based high-contrast instruments \citep{ Hinkley2011,Macintosh2014,Beuzit2019}, allowing the detection of several planetary companions and disks so far \citep[e.g.,][]{Macintosh2015, Chauvin2017, Keppler2018} and large-scale surveys of nearby stars \citep[e.g.,][]{Nielsen2019, Vigan2021}.

However, on top of the complicated diffraction pattern due to telescope segmentation \citep{Yaitskova2003}, the contrast goal of 10\up{-10} translates into tight requirements in terms of observatory stability, with wavefront error budgets on the order of a few hundreds of picometers during the observations \citep{Ultra2019}. To alleviate the constraints on the observatory structure, a solution is to make use of wavefront sensing and control (WFSC) strategies. These methods consist in measuring the wavefront errors with wavefront sensors (WFSs) and correcting for them to obtain the desired shape with one or more deformable mirrors (DMs). In this context, it is possible to distinguish different WFSC applications. One of the main goals is to dig a dark hole \citep[DH,][]{Malbet1995, Borde2006}, an optimized high-contrast region in the image of an observed star to enable exoplanet observations. This approach can also be used to correct for perturbations originating from observatory vibrations and mechanical or thermal drifts, in particular those of the primary mirror segment alignment. These effects lead to a contrast degradation in the coronagraphic image of an observed star, hindering the detection of planets. The wavefront errors cover a wide range of spatial frequencies, from low-order, up to 4-5 cycles per pupil (c/p), to high-order aberrations, at a few tens of c/p. The wavefront errors also cover a scale of temporal frequencies, from low to high range going from the mHz to the kHz regime. Optimal correction requires the combination of simultaneous control loops \citep{Pueyo2019, Guyon2020a, Guyon2020b, Currie2022} to address these errors at different parts of the spatial or temporal frequency ranges.

In the case of a Lyot coronagraph, the focal-plane mask aims to reject the central part of the stellar point-spread function, which corresponds to the fraction of the beam that contains the low-order content of the aberrations. We therefore explore the use of this rejected light to feed a low-order wavefront control (LOWFC) loop that stabilizes the low-order modes of the wavefront. To analyse the low-order information in this rejected light, we use a Zernike Wavefront Sensor \citep[ZWFS,][]{Zernike1934, Bloemhof2003b, Dohlen2004, Wallace2011, N'Diaye2013} to take advantage of its high sensitivity \citep{Guyon2005, Chambouleyron2021b}, the simplicity of its wavefront reconstructor, and the ease of its hardware implementation. In \citet{Pourcelot2022} hereafter \citetalias{Pourcelot2022}, we demonstrated the wavefront stabilisation with a standalone closed loop, stabilizing both the wavefront, and the DH contrast at an average and standard deviation better than 10\up{-7} and 10\up{-8} respectively, under artificially introduced perturbations. While a similar approach has also been demonstrated for the Roman Space Telescope (RST) coronagraphic instrument \citep[CGI,][]{Shi2016, Shi2018, Mennesson2018, Kasdin2020}, enabling deeper contrasts will require the study of the simultaneous use of the ZWFS-based control loop combined with a high-order wavefront control (HOWFC) loop. The algorithms used for the HOWFC in this paper are pair-wise (PW) probing \citep{Borde2006, Giveon2007b, Potier2020} for the electric field estimation in the focal plane and the stroke minimization \citep[SM,][]{Pueyo2009} algorithm to compute the desired DM commands. 

In this work, we investigate a concurrent control loop approach on the High-contrast imager for Complex Aperture Telescopes testbed \citep[HiCAT,][]{hicat1, hicat2, hicat3, hicat4, hicat5, hicat6, hicat7, hicat8} in monochromatic light at the Space Telescope Science Institute (STScI) in Baltimore, USA. In particular, we present an estimation of the interactions between the HOWFC and LOWFC loops, and run it through various configurations. These setups include switching both loops on and off, and two variants of PW probing to estimate the impact of LOWFC on the focal-plane wavefront estimation. In an additional experiment, we limit the maximum speed of the HOWFC loop to allow the LOWFC loop to address more turbulence, and emphasize its positive impact on the DH contrast.

\section{Experimental setup}

\subsection{Optical setup}
We recall the main details of the optical configuration for our study, further details can be found in \citetalias{Pourcelot2022}. We work on stabilizing the contrast with a classical Lyot coronagraph, working with a segmented aperture. The classical Lyot coronagraph is composed of two masks. The first component is an amplitude focal-plane mask that rejects the on-axis light. In our studies, it is implemented as a pinhole in a mirror that lets the rejected light through. The second component, the Lyot Stop, is a diaphragm in the re-imaged pupil plane downstream of the focal-plane mask. Usually implemented with a circular diaphragm, the Lyot stop is slightly undersized with respect to the entrance pupil. The wavefront corrections are performed by using two DMs upstream of the classical Lyot coronagraph, DM1 in a pupil plane and DM2 outside of a pupil plane, for both phase and amplitude control. To address the low-order aberrations, we use a ZWFS in the light rejected by the focal-plane mask to measure them and DM1 to correct for them. Concerning the HOWFC loop, the PW estimation uses several coronagraphic images taken with probes applied on DM1 to estimate the focal-plane electric field, and SM computes commands for both DMs to generate a DH.

\subsection{Implementation on HiCAT}
\begin{figure*}
    \centering
    \includegraphics[width=\linewidth]{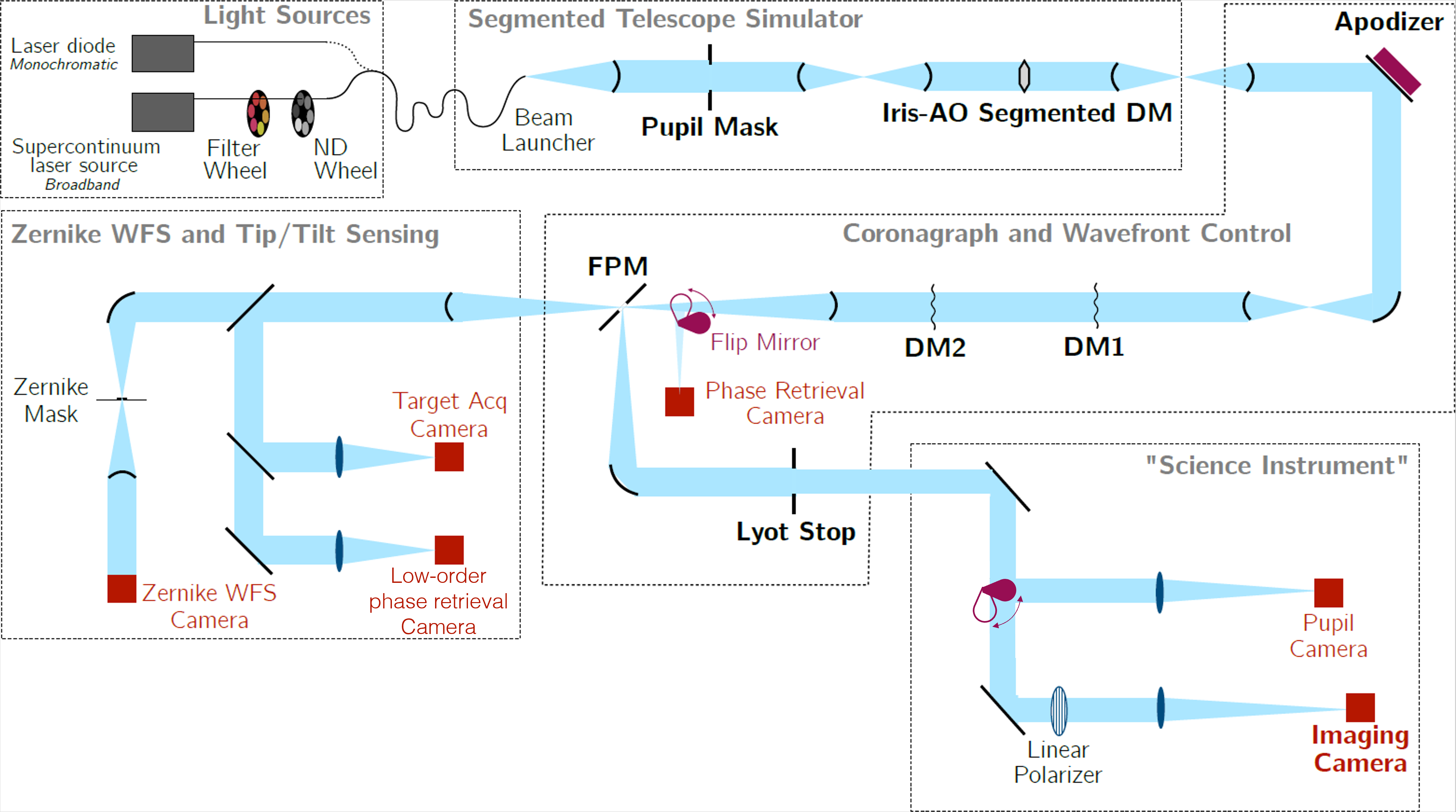}
    \caption{Simplified HiCAT layout in a semi-transmissive representation. The IrisAO segmented DM creates the segemented aperture. The apodizer, in the top-right corner is currently replaced by a flat mirror. The cameras used in this paper are the Zernike WFS camera, in the light transmitted by the focal plane mask (FPM), and the imaging camera in the light reflected by the FPM.}
    \label{fig:hicat_layout0}
\end{figure*}

To test and validate our approach, we use HiCAT in a configuration identical to the one presented in Fig.~3 of \citetalias{Pourcelot2022} and reproduced in Fig.~\ref{fig:hicat_layout0}, with a monochromatic laser source. The relevant values describing the experiment implementation are summarized in Table \ref{tab:hicat_params}. The testbed is composed of several parts. First, there is a telescope simulator that mimics the light from a star through a segmented-aperture telescope. The following arm with the coronagraph includes the continuous DMs, DM1 in a pupil plane and DM2 out of pupil plane, the focal-plane mask, from the Lyot project \citep{Oppenheimer2004}, and the Lyot Stop. The light reflected by the focal-plane mask is sent to the imaging camera, where some light can be picked up with a beamsplitter to re-image the pupil plane. The light rejected by the focal-plane mask is sent to the low-order wavefront sensing (LOWFS) arm with three different detectors: a ZWFS, a target acquisition camera, and a phase retrieval camera. In this work, we focus on the ZWFS as the LOWFS. To isolate it from external perturbations, HiCAT is protected by an enclosure. A flux of dried air at a constant temperature is continuously injected to keep the conditions stable and safe for the DM operations.

\begin{table}[]
    \centering
    \caption{Physical parameter values of the HiCAT testbed used for the experiments.}
    \begin{tabular}{ccc}
        \hline
        \hline
        Parameter & Variable & Value \\
        \hline
        Wavelength & $\lambda$ & 640\,nm \\
        Entrance pupil diameter & $D_{pup}$ & 19.55\,mm \\
        Number of segments & - & 37 \\
        DM 1\&2 actuator count& - & 952 \\
        Total actuator count & $q$ & 1904 \\ \\
        \multirow{2}{*}{focal-plane mask diameter} & \multirow{2}{*}{$d_{\rm{FPM}}$} & $455\,\mu$m \\
        & & $8.52\,\lsd$ \\ \\
        \multirow{2}{*}{ZWFS mask diameter} & \multirow{2}{*}{$d_{Z}$} & $54.3\,\mu$m \\
        & & $1.02\,\lsd$ \\ \\
        ZWFS mask depth & - & 280\, nm \\
        Lyot Stop diameter & $D_{LS}$ & 15\,mm \\
        ZWFS camera model & - &  ZWO ASI178 \\
        ZWFS camera pixel pitch & - & $2.4\,\mu$m \\ 
        ZWFS Region of Interest size & - & 800\,pix \\
        Region of Interest number of pixels & $n$ & 64000 \\
        ZWFS camera max framerate & - & 80\,Hz \\
        Imaging camera model & - &  ZWO ASI178 \\
        \hline
    \end{tabular}
    \label{tab:hicat_params}
\end{table}

\subsection{ZWFS control loop}
\begin{figure*}
    \centering
    \includegraphics[width=\linewidth]{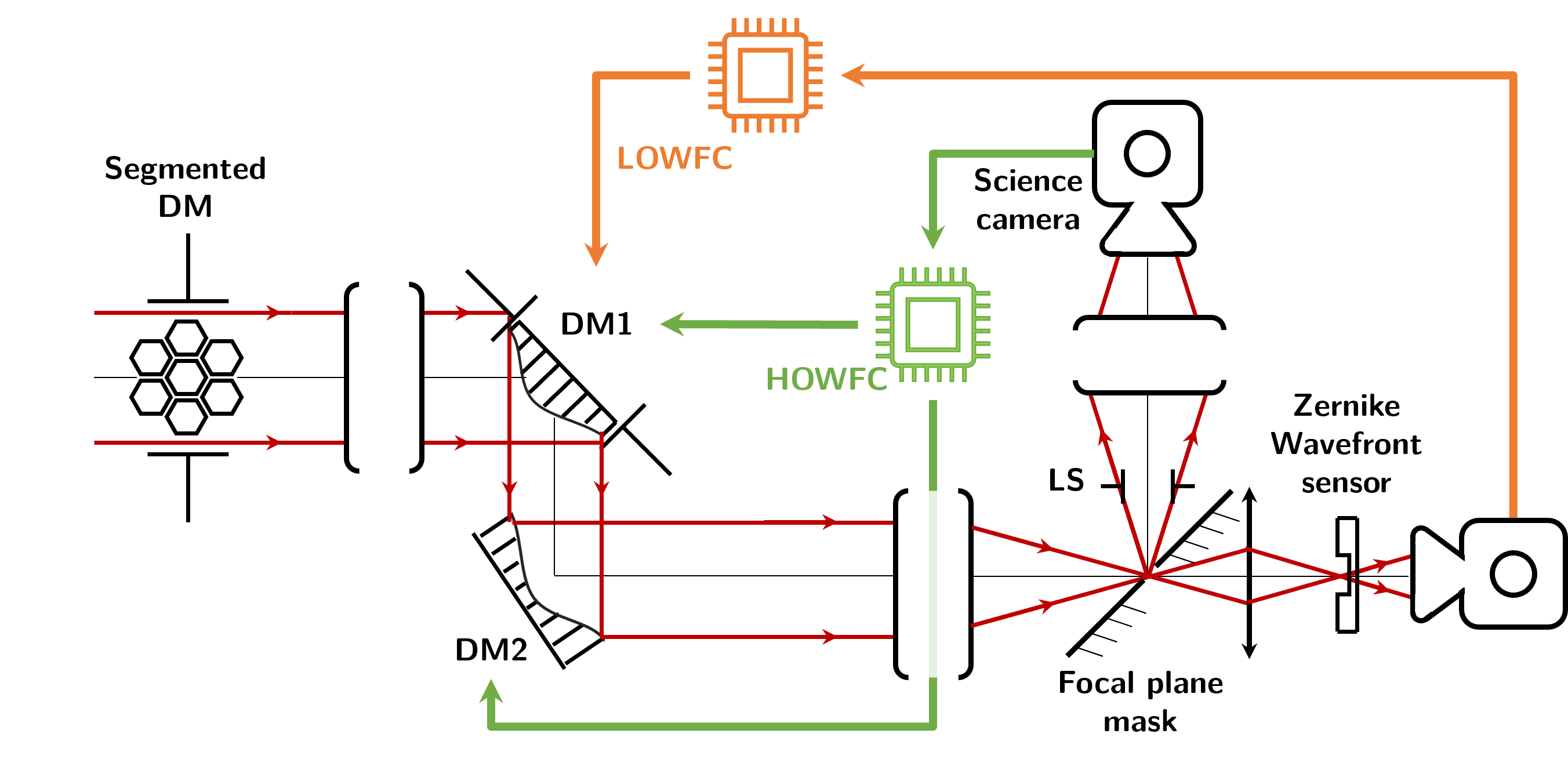}
    \caption{Simplified block diagram of HiCAT with the control loops and their interactions for our experiment. The IrisAO segmented DM creates the segmented aperture. The LOWFC sends commands to DM1 only, while the HOWFC sends commands to both DMs. Loops can be closed or opened during the operations.}
    \label{fig:hicat_layout}
\end{figure*}

Figure~\ref{fig:hicat_layout} represents the block diagram of HiCAT for our experiments and in which the ZWFS control loop is represented in orange. The overall principle remains the same, except that the software implementation of HiCAT has been upgraded since \citetalias{Pourcelot2022} \citep{hicat8, Por2022}. With the use of a service-oriented architecture using shared memory for low-latency inter-process communication, it is now possible to run different testbed operations independently. The calibration method has slightly evolved as well. Instead of poking only 12 Zernike modes, we now consider the first $m$ Zernike modes (excluding piston) following the Noll convention \citep{Noll1976}, with $m = 20$. The Jacobian matrix $\mathbf{J}_Z$ of dimensions $m \times n$ between DM1 and the ZWFS is built using 1000 draws of random combinations of the first $m$ Zernike modes. These draws are poked on DM1, with an expected wavefront error of 10\,nm\,Root-Mean-Square (RMS) per poke. The control matrix $\mathbf{C}$, of dimensions $n \times m$ is then computed by inverting $\mathbf{J}_Z$ using Tikhonov regularization, and a scalar relative regularization parameter of 0.03. The current speed limit imposed on the testbed operations is coming from the maximum frame rate of cameras of 80\,Hz. With this control, the temporal standard deviation for each Zernike mode coefficient is reduced to a few hundreds of picometers RMS. The LOWFC controller is a pure integrator with a gain of 0.01, at a loop speed of 80\,Hz, limited by the camera readout speed. 

\subsection{Dark hole digging with stroke minimization}
DH digging usually relies on two steps: the measurement of the electric field in the DH region in the focal plane and the computation of the DM commands that will efficiently remove the light in the DH. HiCAT currently relies on PW for the determination of the electric field in the focal plane. PW probing applies a set of DM commands, called probes, that modulate the energy distribution on the imaging camera, and especially in the DH area. By using probes with enough diversity, a testbed model and by solving a linear problem between the DM commands and the focal-plane electric field, an estimation of both the amplitude and the phase of the electric field is obtained in the focal plane. The choice of the probes is made by considering several parameters, such as the area in the science image we want to control. Two sets of probes are currently used on HiCAT for PW performance studies. The most basic probes are made by poking single DM actuators, hereafter single-actuator probes. The second set called ``HiCAT probes'' can be computed by solving an optimization problem to find the DM command that produces a specific modulation of the electric field in the DH \citep{Will2021}.

\begin{figure}[h!]
    \centering
    \includegraphics[width=\columnwidth]{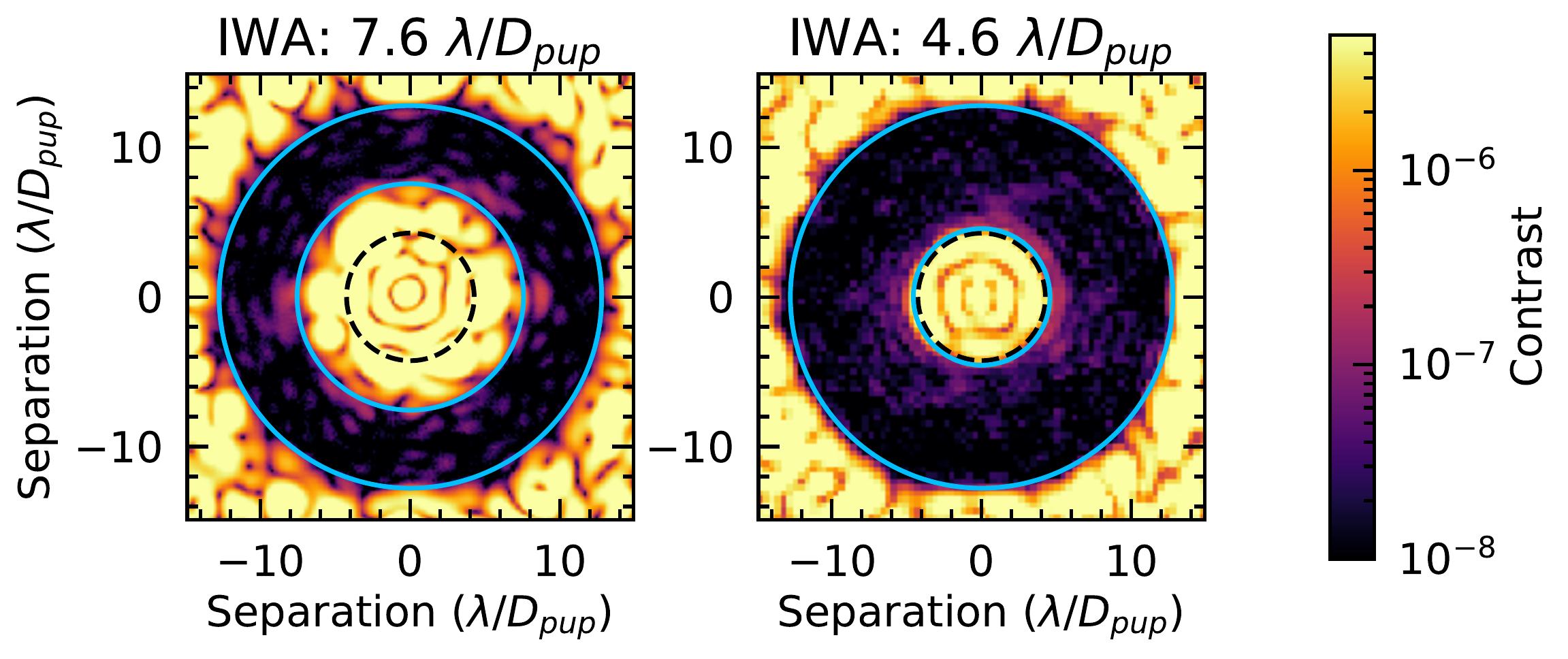}
    \caption{DH examples with an outer working angle of 12.8\,\lsd, with an IWA of 7.6\,\lsd (left) and 4.6\,\lsd (right). The blue circles define the edges of the controlled DH. The black dashed circle delimits the focal-plane mask area. The residual DH speckles are removed with a faster HOWFC loop thanks to the new HiCAT software architecture.}
    \label{fig:two_dhs}
\end{figure}

By using the PW estimation and a DM-to-focal-plane relation such as a Jacobian matrix built between the DMs and the focal-plane electric field, it is possible to perform an optimization of the DM commands to minimize the energy in the DH. Several algorithms exist, among which SM \citep{Pueyo2009, Mazoyer2018a, Mazoyer2018b} that is implemented on HiCAT. An alternative algorithm is electric field conjugation \citep[EFC,][]{Giveon2007a} that aims at solving a similar problem with a different energy minimization strategy. If the electric field conjugation algorithm requires less computation time and therefore runs faster than SM, it usually provides larger actuator strokes on the DMs. Alternative strategies have successfully been implemented and tested on HiCAT \citep[e.g.,][]{Will2021} but they are not considered here. Fig.~\ref{fig:two_dhs} shows an example of DHs generated with SM during HiCAT experiments with an outer working angle of 12.8\,\lsd and two different inner working angles (IWA). While the contrast of $~2\times10^{-8}$ could only be reached with a large IWA of 7.6\,\lsd in \citetalias{Pourcelot2022}, the new, fast software architecture allows for a loop frequency gain of a factor of 10 and therefore a better compensation of the turbulence in the HiCAT environment. This enables the digging of a DH with the same contrast like previously but with a reduced IWA down to $4.6\,\lsd$. This is the value set for the IWA of the DH used in the rest of the paper. 

\section{Interactions between the loops} \label{sec:interactions}

When running two wavefront control loops in parallel, we want to avoid any cross-talk between the processes. As the experiment goes on, SM is going to refine the DM commands to reduce the intensity in the DH. From the ZWFS point of view, it means the application of a new DM offset at each iteration. The sensitivity loss of the ZWFS due to this offset change has been addressed in \citetalias{Pourcelot2022} in the case of a large IWA of $7.6\,\lsd$, and a similar behavior is observed with an IWA at $4.6\,\lsd$. Overall, this offset change does not prevent the ZWFS control loop to efficiently measure the low-order aberrations. 

Conversely, we want to avoid the ZWFS control loop to correct for SM updates or for the probes introduced by PW. The success of this will depend on the response measurements of these commands by the ZWFS. In our configuration with contrast levels of 10\up{-8}, we estimate the projection of the PW probes or SM commands on the modes controlled by the LOWFC loop using the following procedure. For a given command $\mathbf{v}$ controlling both DMs, of dimensions $1 \times q$, we project $\mathbf{v}$ on the Zernike polynomial basis that is used to compute $\mathbf{J}_Z$, leading to the term $\mathbf{v}_{//}$ of dimensions $1\times m$. As $\mathbf{J}_Z$ is a model of our system, we can therefore estimate the response on the ZWFS detector $\mathbf{b}_{img}$ of size $n$ by computing $\mathbf{b}_{img} = \mathbf{J}_Z\mathbf{v}_{//}$. Finally, by using the control matrix $\mathbf{C}$, the pseudo-inverse of the Jacobian matrix $\mathbf{J}$, we can assess the command $\hat{\mathbf{v}}$ sent by the ZWFS control loop to DM1 to correct for the perturbation induced by $\mathbf{v}$ before applying the control gain. This estimation can also be used to avoid the correction of SM commands and PW probes by the ZWFS, as showed later in Sec.~\ref{sec:slow_loops}.

\begin{figure*}[h!]
    \centering
    \includegraphics[width=\linewidth]{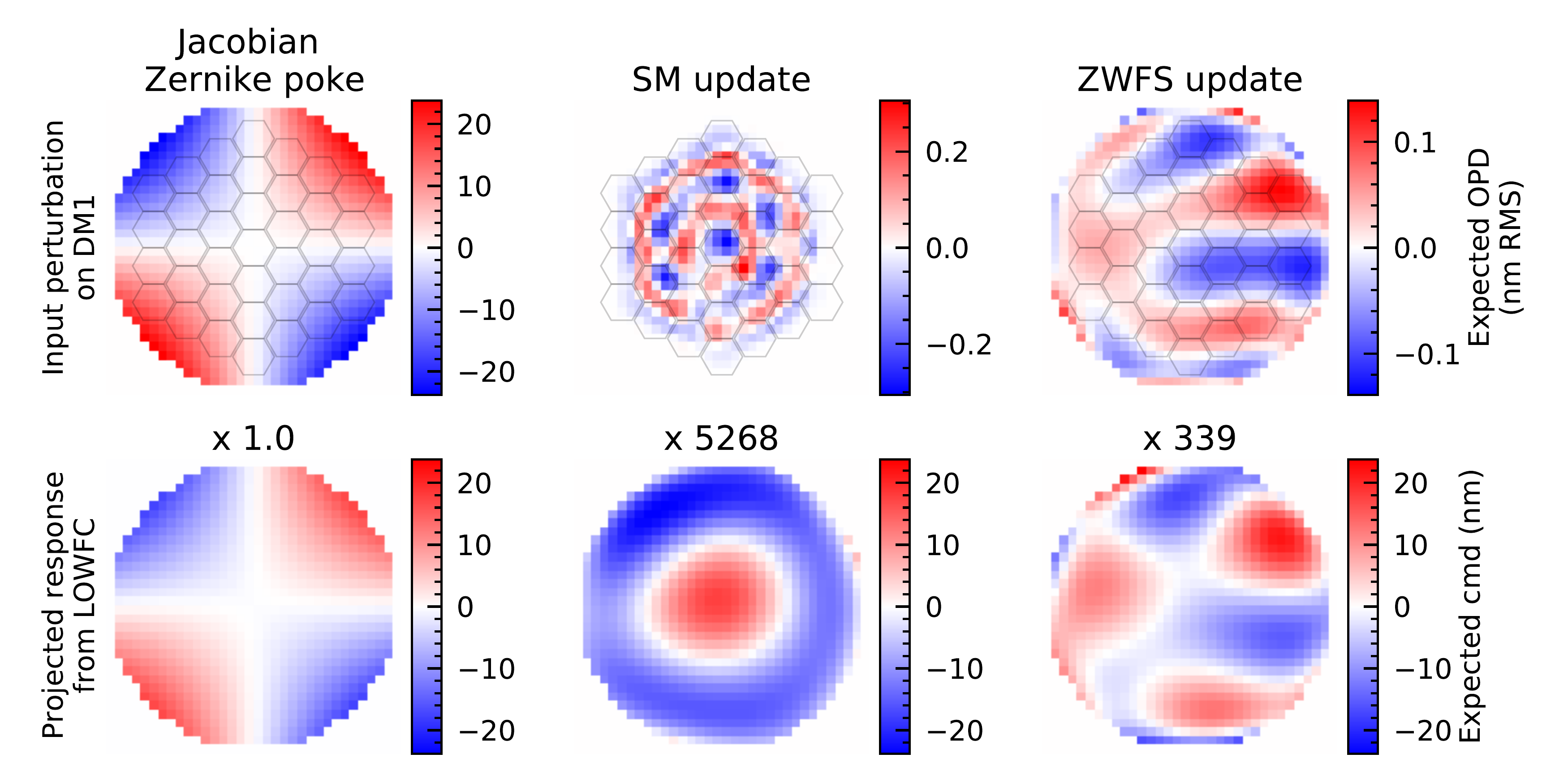}
    \caption{Examples of different perturbations introduced on DM1 (top) and their corresponding LOWFC loop responses on DM1 (bottom), obtained by projecting the command on the controlled basis of the LOWFC loop and using the control matrix to estimate the feedback command. From left to right: the Zernike mode $Z_4$ used for the interaction matrix calibration, a single update from SM after convergence of the DH digging, and a command update from the LOWFC loop to correct for residual turbulence in closed loop. In the bottom row, all the displayed commands are scaled to fit the same color bar. From left to right, the commands have been multiplied by 1, 5268 and 339 respectively to be displayed with the same color bar, and the respectively applied amplification factor is displayed on top of each command. The commands shown here expand over the whole controllable area of DM1. The hexagons represent the projection of the segmented aperture onto DM1.}
    \label{fig:impact_examples}
\end{figure*}

These estimations are given in Fig.~\ref{fig:impact_examples} for a pure Zernike mode, a single SM update after convergence of the DH contrast, and a correction for a single perturbation measured by the ZWFS while in closed loop. For all the estimated commands, we use the same colorbar and we display the amplification factor applied to each command in the plot to match the colorbar scale. In this configuration, the SM update triggers a faint response of the LOWFC loop. This result is expected as the SM command contains small DM strokes and it is computed to modulate the DH intensity. Since the DH reaches in to the edge of the focal-plane mask, its corresponding spatial frequencies differ from the spatial frequencies that are seen by the ZWFS. In comparison with the SM commands, the command sent by the LOWFC loop to correct for turbulence is 10 times fainter in peak-to-valley but triggers a response with a peak-to-valley value which is 20 times larger than the SM command. In practice, we expect the command triggered by the ZWFS for a SM update to be even smaller because we use two DMs on HiCAT to correct for phase and amplitude. In the case of a classical Lyot coronagraph with a segmented aperture, the amplitude correction dominates over the phase correction due to the segment gaps in our pupil. DM2 will be used to correct for these amplitude errors, introducing both the desired amplitude correction and undesired phase correction which is compensated by DM1 \citep{Mazoyer2018b}. 

\begin{figure*}[h!]
    \centering
    \includegraphics[width=\linewidth]{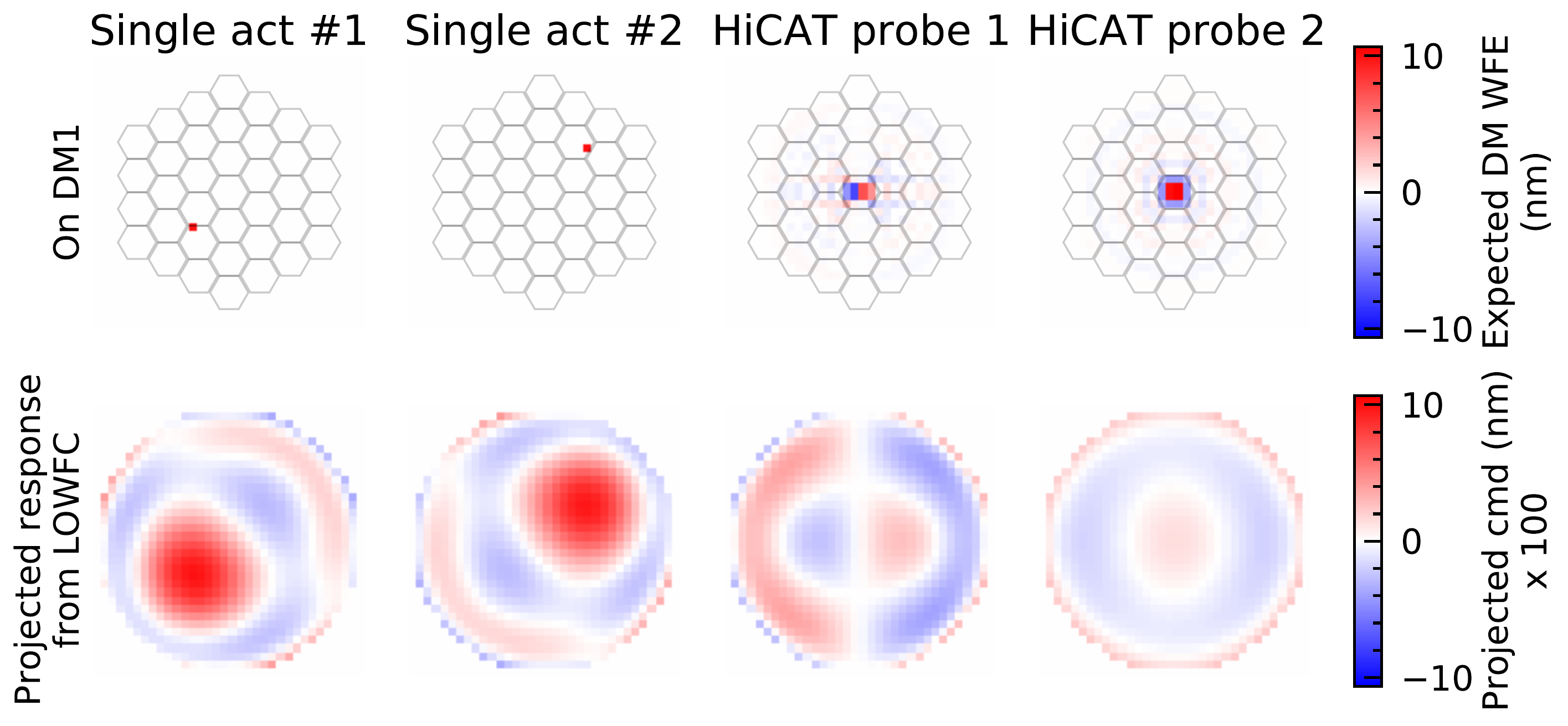}
    \caption{Examples of different PW  probes introduced on DM1 (top) and their corresponding LOWFC loop responses on DM1 (bottom) obtained by projecting the command on the controlled basis of the LOWFC loop and using the control matrix to estimate the feedback command. From left to right: two examples of single-actuator probes poked by 10\,nm in wavefront error; two of the 4 HiCAT probes designed to specifically modulate the electric field in the DH. To match the top and bottom row color bars, the bottom commands have been scaled by a factor of 100. The commands shown here are over the whole controllable area of the DM1. The hexagons represent the projection of the segmented aperture onto DM1.}
    \label{fig:probes_impact}
\end{figure*}

The other conflicting commands with the ZWFS arise from the probes introduced by PW. Following the same principle as for Fig.~\ref{fig:impact_examples}, Fig.~\ref{fig:probes_impact} shows examples of commands triggered by the ZWFS loop when single-actuator and HiCAT probes are sent to DM1. To match the plots with the same color bars, the DM response have been amplified by a factor of 100. Overall, these responses are also small, with the same order of magnitude as the single ZWFS turbulence correction. Similarly to the SM commands, the HiCAT probes are less seen by the ZWFS than the single-actuator probes: they are optimized to modulate the DH intensity, which means they are targeting wavefront error spatial frequencies beyond the ZWFS range. 

Overall, the commands introduced by SM and PW present a limited impact on a single ZWFS control-loop command. Furthermore, on HiCAT, the PW estimation runs at 80\,Hz which is close to the frequency of the LOWFC loop, and SM at 0.4\,Hz with an iteration every 2.5\,s. This translates into approximately 200 LOWFC loop iterations for one SM command. With a control gain of 0.01 and considering the ZWFS amplification factors in Figs.~\ref{fig:impact_examples} and~\ref{fig:probes_impact}, the ZWFS control loop is therefore unlikely to disturb the probes and the SM updates by more than 0.01\% and 0.1\%, respectively. As a result, we choose to disregard these perturbations for the experiments shown in Sec.~\ref{sec:fast_results} with a 10\up{-8} contrast regime. Further analysis of the PW estimation degradation due to LOWFC would be required for deeper contrasts but this is beyond the scope of this paper and will be addressed in future work.  

\section{Parallel operations of low-order and high-order wavefront control loops}
\subsection{Control loops architecture}
\begin{figure}
    \centering
    \includegraphics[width=\columnwidth]{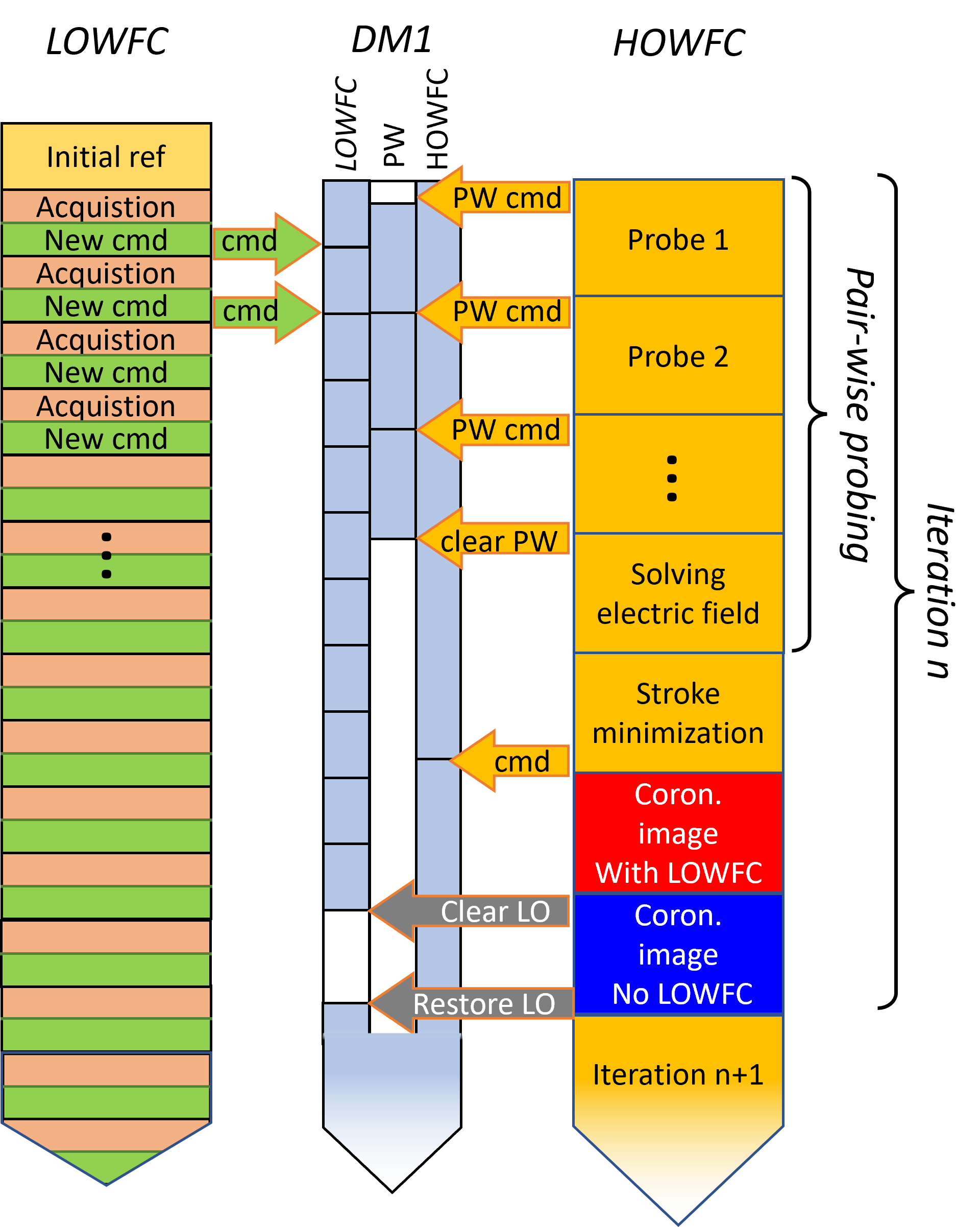}
    \caption{Vertical timeline of the parallel operations on HiCAT with three processes. The DMs are operated by a process that reads from all the different channels (LOWFC, PW and HOWFC here) and apply the shape corresponding to the sum of the different contributions. The LOWFC process starts by taking an initial reference, and then reads images from the camera and computes corrections sent to the LOWFC DM channel. The HOWFC reads runs PW probing, using a dedicated DM channel, and computes DM corrections with SM for the HOWFC channel. It can also interrupt the LOWFC by clearing the channel for specific acquisitions without LOWFC. Each process runs independently from the other. Data is transferred from cameras or to DMs through shared memory.}
    \label{fig:parallel_loops}
\end{figure}

In this section, we present the results of simultaneous operations of the DH digging process with PW and SM, and ZWFS control. Considering the conclusions of Sec.~\ref{sec:interactions}, we run the loops in a completely asynchronous manner. The principle of these operations is detailed in Fig.~\ref{fig:parallel_loops}, where both loops run in a separate process and send commands to a virtual DM channel. Another process then sums the commands sent to the channels and applies the final command to the DMs. Two configurations are explored for PW: using single-actuator or HiCAT probes. For both of these experiments, the protocol is the same: we pre-compute a DH solution using a fast electric field conjugation algorithm that manages to provide DM solutions yielding an averaged contrast smaller than $5\times10^{-8}$ in the DH. The experiment then goes through 5 parts: (0) SM runs alone; (1) SM runs in parallel with ZWFS; (2) SM runs alone again; (3) SM runs in parallel with ZWFS again; (4) SM is stopped while ZWFS is running. During (4), for each contrast measurement, another measurement is performed with the ZWFS loop opened and DM command reset to the first command of (4). All along the experiments, a measurement of the direct flux without the focal-plane mask is performed every 10 iterations to get an accurate contrast normalization. 

\begin{figure}
    \centering
    \includegraphics[width=\linewidth]{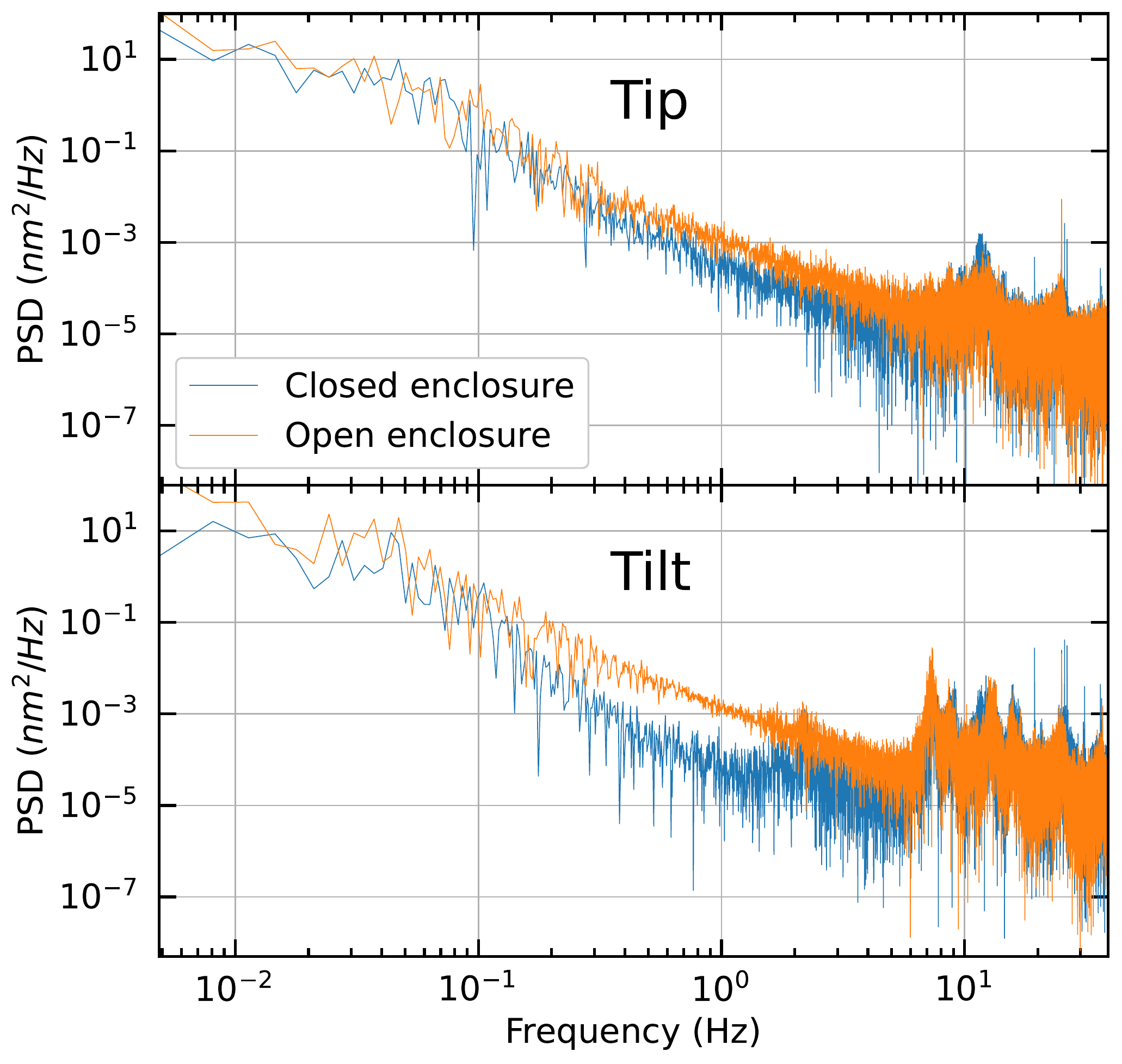}
    \caption{Examples of power spectral density (PSD) of the tip (top) and tilt (bottom) perturbations at the level of the ZWFS on HiCAT. For both modes, a measurement was performed with all the panels closed (in blue) and with an open panel of the enclosure and a high dry air flow (orange).}
    \label{fig:psd}
\end{figure}

Recent studies of the testbed stability have identified a warm flip-mount motor generating the air turbulence that was characterized in \citetalias{Pourcelot2022}. After moving this component, the beam is now more stable without any noticeable contrast drift during experiments running longer than 1\,h, even without control and with contrasts around $3\times10^{-8}$. To study the behavior of the ZWFS control loop with DH digging in a less favorable environment, these two experiments are performed with a $10\times50\,\mathrm{cm^2}$ opening in the HiCAT enclosure, creating extra turbulence within the testbed. In addition, to keep the DMs safe during the operations, the dry air flow injected in the bench has been increased to compensate the influx of humid air through the opening in the enclosure and stabilize the environment humidity and temperature in the turbulent airflow on testbed. Examples of power spectral density of the turbulence for the tip and tilt modes is presented in Fig.~\ref{fig:psd}. They show a turbulent motion in the low frequency range below 1\,Hz that we associate to turbulence. Perturbations at higher frequencies around 10\,Hz are also present and are associated to vibrations of the testbed. They are not specifically linked to the generated turbulence. The behavior is very similar with the other Zernike modes but almost two orders of magnitude smaller.

\subsection{Results} \label{sec:fast_results}

\begin{figure*}[h!]
    \centering
    \includegraphics[width=\linewidth]{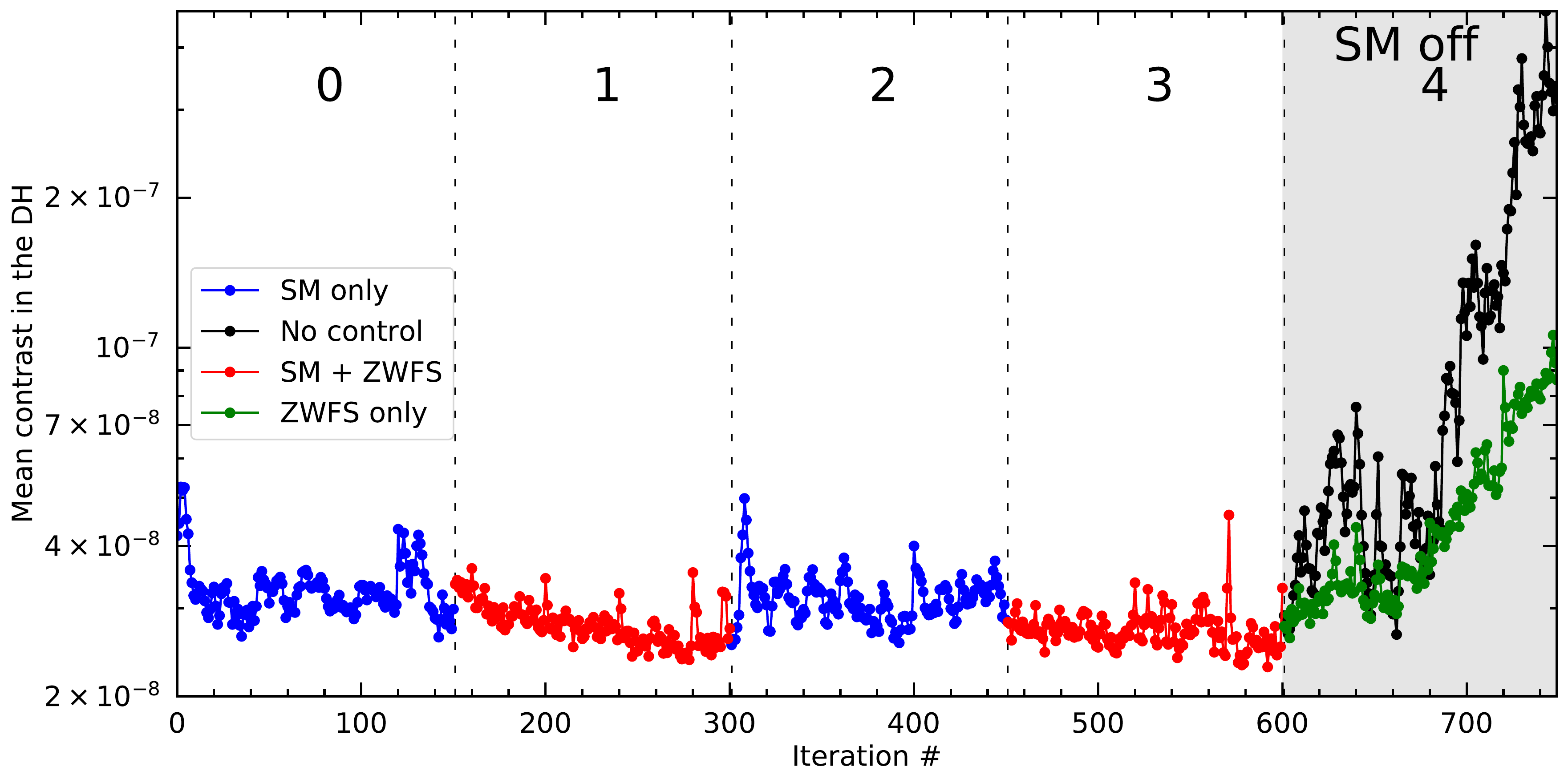}
    \caption{Mean DH contrast while PW with single-actuator probes and SM are running, without LOWFC in blue in section (0) and (2), and with it in red in sections (1) and (3). SM is turned off in section (4). In (4), the mean contrast is measured while LOWFC is running. Once per contrast measurement, the LOWFC is interrupted and reset to its initial command in (4) to get an open-loop measurement, plotted in black. When on, the HOWFC loop runs at 0.4\,Hz and the LOWFC loop at 80\,Hz.}
    \label{fig:cont_plot_sap}
\end{figure*}

\begin{figure*}[h!]
    \centering
    \includegraphics[width=\linewidth]{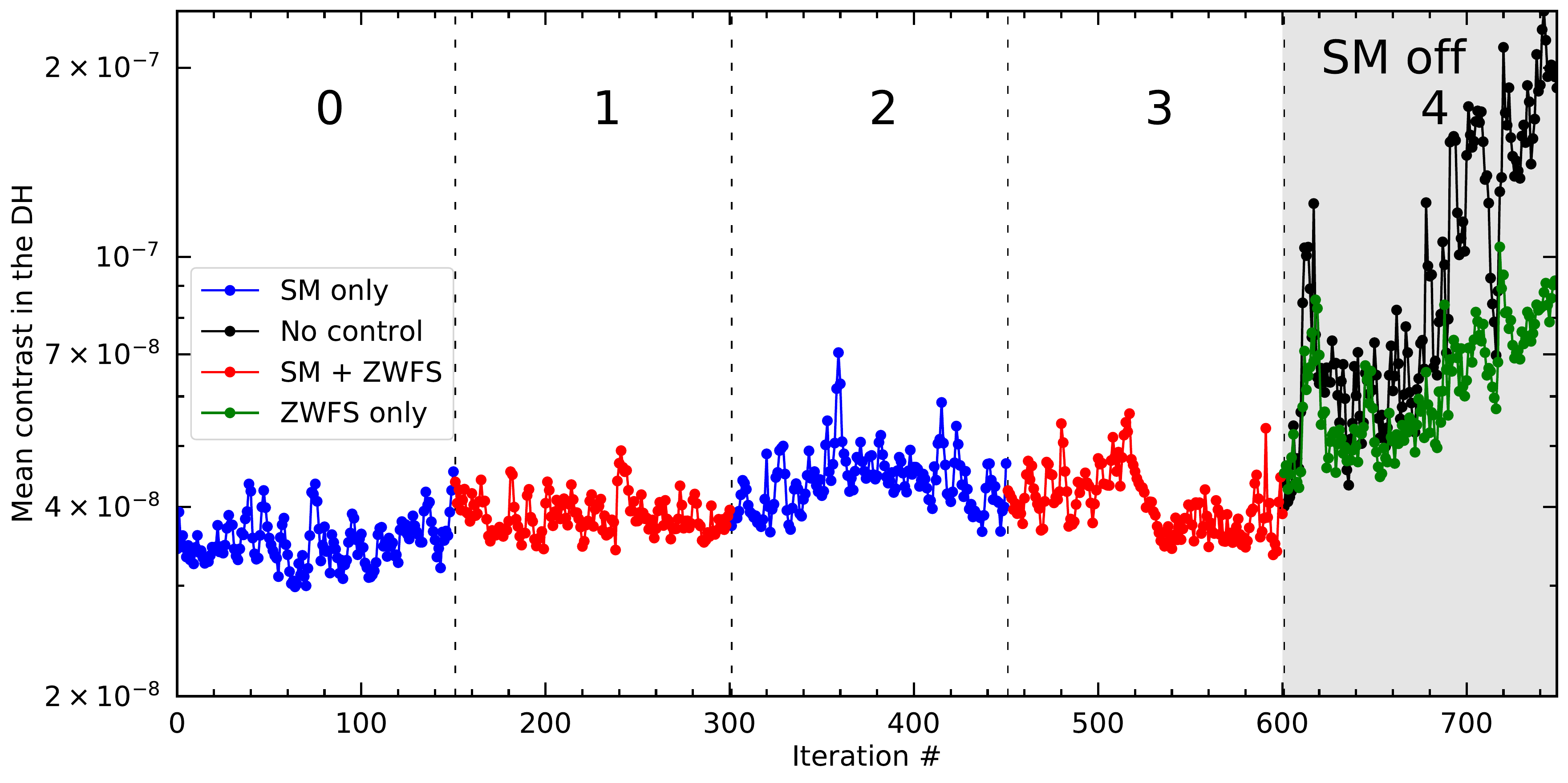}
    \caption{Same as Fig.~\ref{fig:cont_plot_sap} but using PW with HiCAT probes instead of single-actuator probes.}
    \label{fig:cont_plot_hcp}
\end{figure*}

\begin{figure*}
    \centering
    \includegraphics[width=\linewidth]{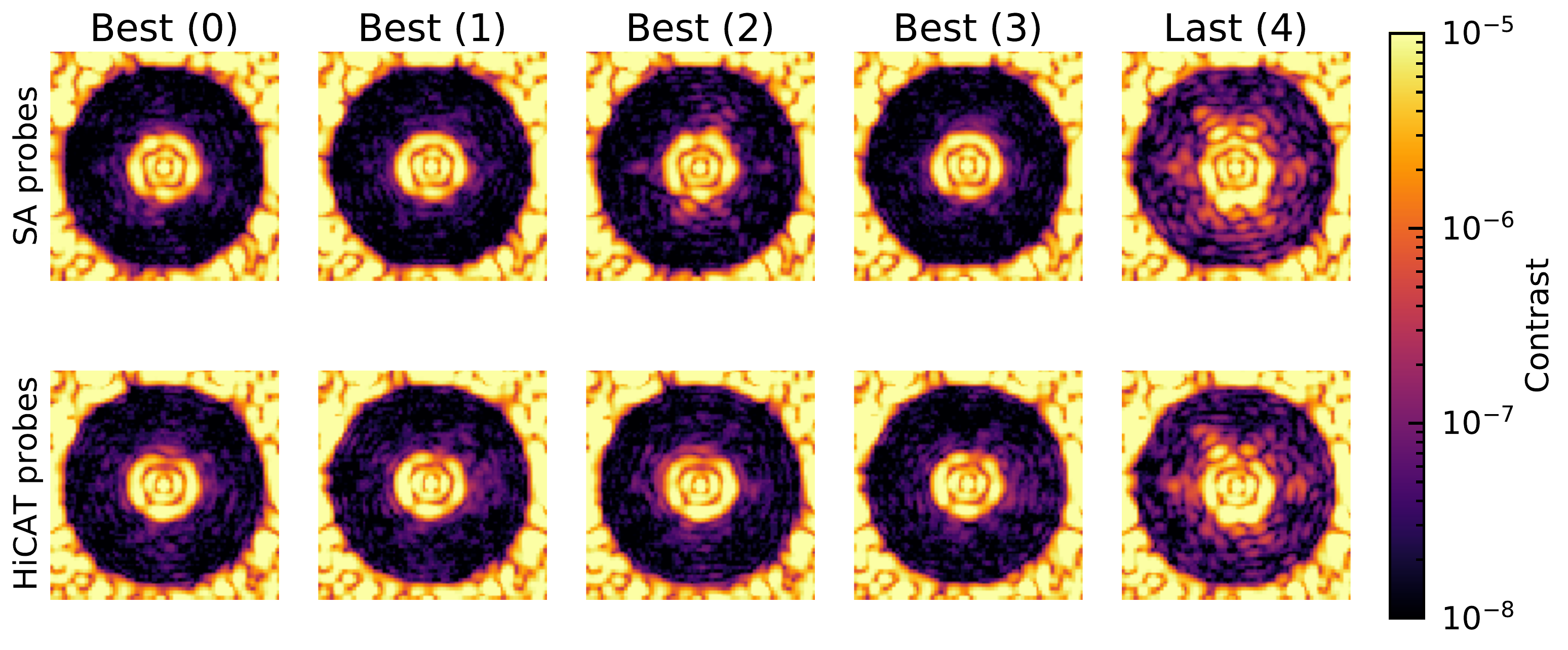}
    \caption{Examples of focal-plane DH images for the experiment with single-actuator probes (top) and HiCAT probes (bottom). For intervals (0) to (3), this corresponds to the image yielding the best mean spatial contrast in the respective operational sections in Figs.~\ref{fig:cont_plot_sap} and \ref{fig:cont_plot_hcp}. The image for interval (4) is the last one from the open-loop data, shown in black in both figures. }
    \label{fig:dh_evolution}
\end{figure*}

The mean DH contrast is presented as a function of the iterations in Figs.~\ref{fig:cont_plot_sap} and \ref{fig:cont_plot_hcp} for the single-actuator and HiCAT probe scenarios. Examples of DH images extracted from the two experiments are given in Fig.~\ref{fig:dh_evolution}, with the image yielding the best DH contrast for intervals (0) to (3) and the respective last images for intervals (4). The corresponding contrast statistics are detailed in Table~\ref{tab:dh_stab_cont}. For both probe types, the HOWFC loop manages to correct for most of the wavefront errors, yielding a fairly stable contrast in the intervals (0) to (3), especially in the intervals (0) and (2) where no LOWFC loop is running. Conversely, large drifts are visible in (4), where we observe a contrast loss of an order of magnitude compared to sections (0)-(3). Consistently with the previous blind DH stabilization \citepalias{Pourcelot2022}, the LOWFC loop slows down the contrast drift. From the difference between intervals (0) and (2), and intervals (1) and (3), we draw several conclusions. 

\begin{table*}[]
	\centering
	\caption{Mean DH contrast statistics over the different intervals for the single-actuator probes experiment and HiCAT probes experiment.}
	\begin{tabular}{cccccccc}
		\hline
		\hline
		Situation & Stat & 0 & 1 & 2 & 3 & 4 with ZWFS &4 no ZWFS  \\ 
		 \hline 
        ZWFS & & off & on & off & on & on & off \\
        SM & & on & on & on & on & off & off \\ 
        \hline
		\multirow{2}{*}{Single-actuator probes}&Mean & 4.5e-08 & 3.6e-08 & 4.0e-08 & 3.4e-08 & 5.4e-08 & 1.4e-07\\
				& $\sigma_t$ & 6.0e-09 & 3.6e-09 & 4.7e-09 & 3.3e-09 & 2.1e-08 & 1.3e-07 \\		\hline
		\multirow{2}{*}{HiCAT probes}&Mean & 4.5e-08 & 4.7e-08 & 5.3e-08 & 4.8e-08 & 7.0e-08 & 1.3e-07\\
				& $\sigma_t$ & 4.7e-09 & 3.6e-09 & 6.7e-09 & 5.9e-09 & 1.5e-08 & 7.3e-08 \\		\hline
	\end{tabular}
	\label{tab:dh_stab_cont}
\end{table*}

First, we show a HOWFC and LOWFC loop combination that manages to keep the DH stable at a contrast of $\sim 5\times10^{-8}$ under large natural drifts. Concerning the standard deviation $\sigma_t$, it is typically maintained at an order of magnitude lower than the averaged contrast, and with $\sigma_t = 5\times10^{-9}$ over 25\,min of an experiment. This asynchronous implementation does not introduce conflicts between the two controls. At slightly better contrasts, the ``Très Haute Dynamique 2'' testbed is routinely operating a fast low-order correction loop \citep{Galicher2020}. However, their approach uses the light reflected by the Lyot stop and it addresses only the control of tip and tilt on a separate steering mirror. Contrasts deeper than 10\up{-8} have also been obtained by RST/CGI \citep{Zhou2019, Zhou2020} with a HOWFC loop including a LOWFC loop \citep{Shi2016, Shi2017, Shi2018}, but our approach differs in several aspects. (i) While RST relies on a monolithic primary mirror, HiCAT implements a segmented aperture with an IrisAO deformable mirror. (ii) Although the tests on RST have been performed with a fast LOWFC loop at frequencies up to 1\,kHz, this loop only addresses tip/tilt ($Z_2$, $Z_3$) at this speed with a steering mirror to control them. The other modes are controlled at much slower rates, around 0.2\,Hz for defocus ($Z_4$) with their DM2 and 5\,mHz for other modes \citep{Cady2017, Shi2017, Seo2017, Seo2018} with their DM1. At the same time, the HOWFC loop on RST is running on both DMs at around 0.1\,Hz, limiting the possible cross-talks with LOWFC. The RST tests were also performed on a testbed located in a vacuum, where it is only perturbed by known artificial drifts with a few Zernike modes while we operate our testbed in air and under strong turbulence with an opened enclosure. (iii) The RST/CGI observation mode is designed to dig a DH on a bright star and then slew to the science target, using the LOWFC loop only to correct for perturbations. In our case, we aim at demonstrating the continuous use of both loops while observing. 

Second, the contrast gain in both temporal average and standard deviation from Table~\ref{tab:dh_stab_cont} are moderate but noticeable, especially in the single-actuator probe experiment, reaching a mean contrast of $3.4\times10^{-8}$ and a value of $\sigma_t$ of $3.3\times10^{-9}$ over the interval (3). While we expect the contrast with LOWFC to be better because of the turbulence that introduces a lot of low-order aberrations, the new implementation of the HOWFC control loop runs fast enough to handle a large part of the low-order corrections. From our experience with electric field conjugation on HiCAT, digging a DH with faster iterations allows even deeper contrasts in the presence of turbulence. Therefore, the contrast we observe is very likely to be driven by the internal turbulence that evolves faster than what the HOWFC loop can handle in the high-order modes. Further experiments are required to fully understand the limitations in these conditions. 

\begin{figure}
    \centering
    \includegraphics[width=\linewidth]{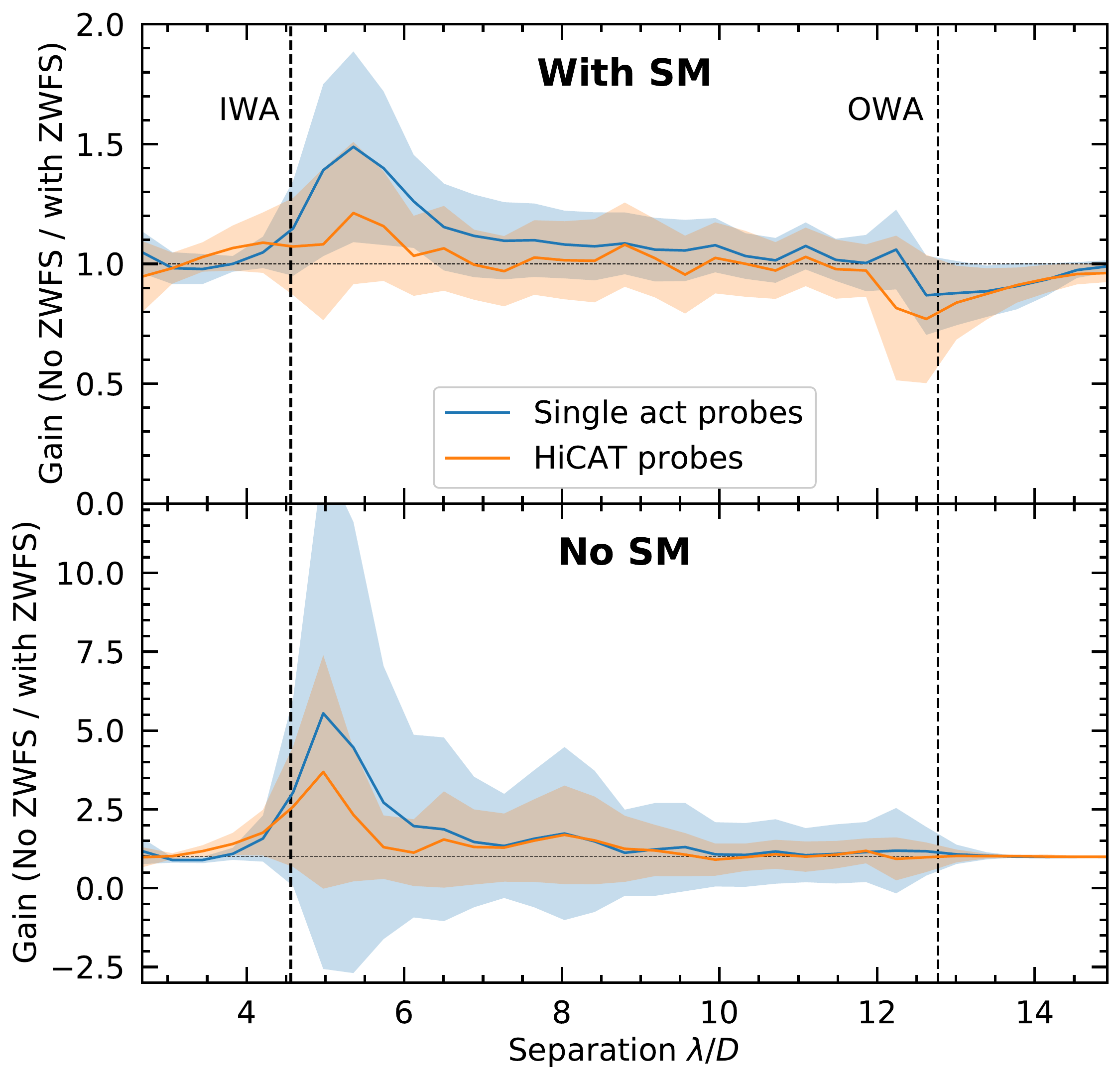}
    \caption{Contrast gain due to ZWFS as a function of angular separation in the DH, for experiments with single-actuator and HiCAT probes. Solid lines represent the ratio of the contrast azimutal average between the cases without and with LOWFC. The filling colors show the standard deviation of the measurement. The dashed horizontal lines are drawn at a gain of 1 in both plots. The top plot represents the values of the intervals (0) and (2) over the values of the intervals (1) and (3). The bottom plot corresponds to the values of the interval (4). The gain lower than 1 at the OWA is due to edge effects in the numerical computation of the contrast.}
    \label{fig:gain_vs_separation}
\end{figure}

Third, even if the temporal contrast averages are moderate, the ratio of contrast between the cases without and with LOWFC as a function of angular separation in the focal plane emphasizes the action of the loop. The corresponding curves are displayed in Fig.~\ref{fig:gain_vs_separation}. For each probe, the plots show the ratio between the azimuthally averaged contrast over sections (0) and (2), and over sections (1) and (3). The maximum contrast improvement by a factor of up to 1.5 is located at separations around $4.6\,\lsd$, as expected - this corresponds to the edge of the focal-plane mask, and therefore the theoretical sensing limit of the ZWFS through the Lyot mask. Since the LOWFC loop controls a limited number of 20 Zernike modes here, its impact is negligible at larger separations. The same plot is also displayed for section (4) in Fig.~\ref{fig:gain_vs_separation} bottom plot, showing the same behavior, but with a much larger gain up, to a factor of 5. This is consistent with the first point above: in this experiment, both loops control overlapping spatial frequencies of the wavefront error without cross-talk. The gains at larger separations are negligible here as well. 

Finally, the experiments have slightly better results with single-actuator probes, even though their interaction with the LOWFC loop was supposed to be greater with the HiCAT probes. However, these single-shot experiments do not allow us to conclude on any significant difference between the two types of probes, the different behaviors in Fig.~\ref{fig:cont_plot_sap} and~\ref{fig:cont_plot_hcp} being possibly due to external factors such as turbulence. The real limitation due to loop interactions will probably appear when working at deeper contrasts. We are still investigating the exact limitations on HiCAT of the electric field estimation by PW. These interactions could degrade wavefront estimation that would result in reaching a contrast floor at some point. Therefore, using probes as invisible as possible to the ZWFS could prove of interest in the future.

\subsection{Experiment with slower high-order wavefront control loop} \label{sec:slow_loops}
The gain provided by the ZWFS depends not only on the input turbulence, but also on the amount of low-order aberrations the HOWFC loop corrects for. By degrading the correction performed by the high-order controller we can emphasize the improvement due to the parallel use of the ZWFS. This also degrades the contrast performance by an order of magnitude, and is not representative of the current operations of HiCAT.

For this goal, we use HiCAT in a different setup, with a HOWFC loop running 8 times slower, at 0.05\,Hz and the LOWFC loop around 2.5 times slower at 30\,Hz. In this experiment, the HiCAT probes are used for PW. With the slower HOWFC and similar air turbulence, the contrast in the DH is stuck above 10\up{-7}. The speed difference between the two controllers now being much larger, we expect the cross-talk between them to be more challenging. For this experiment, we have implemented a communication between the loops to offload the HOWFC loop commands to the LOWFC reference: each time PW or SM sends a command on the DMs, the command is sent to the LOWFC loop as well. The command is then projected on the controlled modes of the LOWFC loop and multiplied by the Jacobian matrix $\mathbf{J}_Z$ to estimate the corresponding variations on the ZWFS signal. While there are many ways for improvement here, our code modifications for synchronisation add overheads in the computation and slow down the LOWFC loop from 80\,Hz to 30\,Hz. As a result we increase its gain from 0.01 to 0.15, a value that empirically proved to be efficient. Considering the loop frequencies, the LOWFC loop produces around 600 corrections in a single HOWFC loop iteration, while with the previous setup it was only around 200. 

\begin{figure*}
    \centering
    \includegraphics[width=\linewidth]{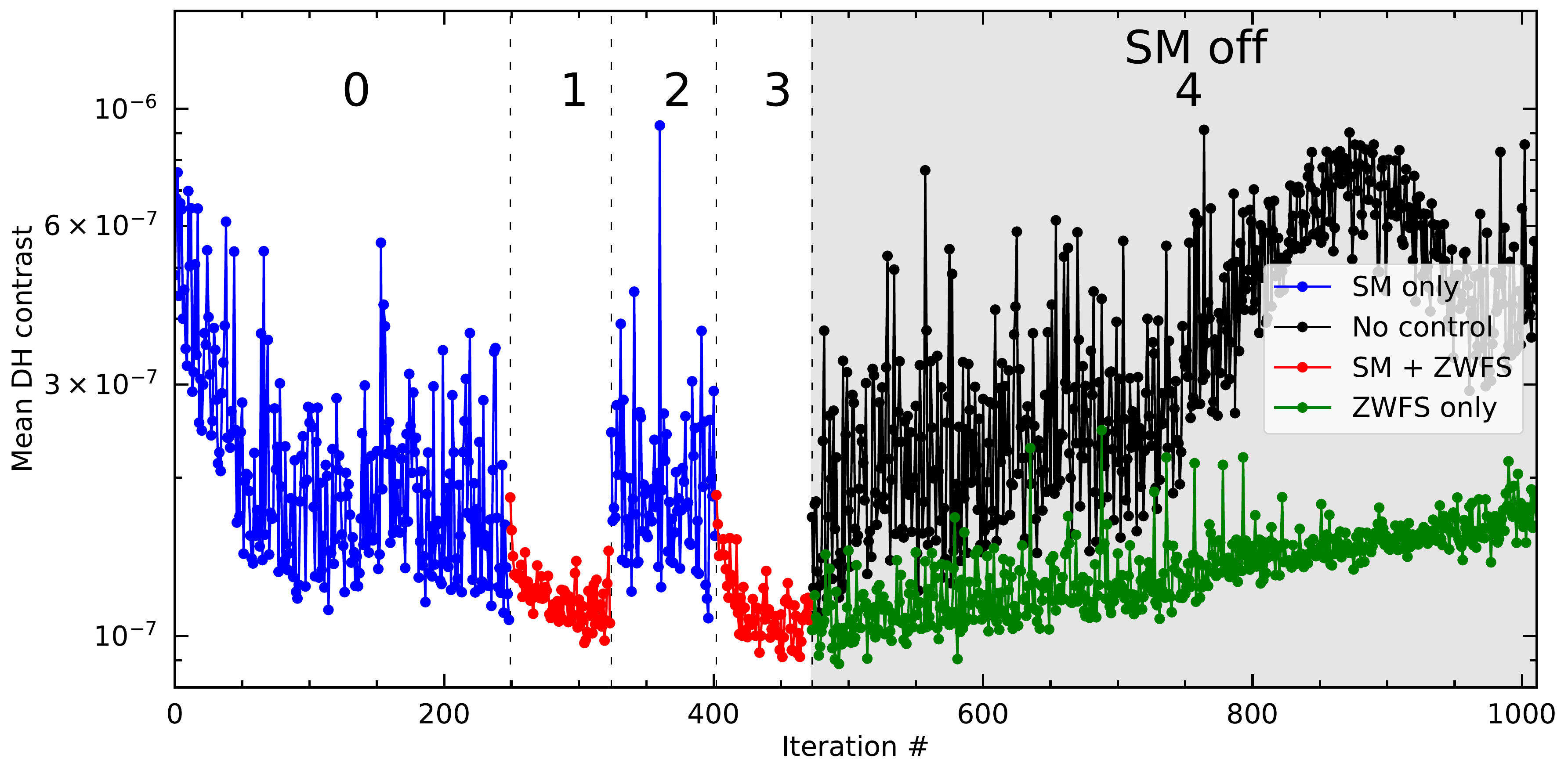}
    \caption{Evolution of the mean contrast in the DH during HOWFS (using SM) as a function of iterations during different intervals: without LOWFC in blue in section (0) and (2), and with it in red in sections (1) and (3). SM is turned off in section (4). In (4), the mean contrast is measured while LOWFC loop is running (green). Once per contrast measurement, the LOWFC loop is interrupted and reset to its initial command in (4) to get an open-loop measurement, in black. When on, the HOWFC loop runs at 0.05\,Hz and the LOWFC loop at 30\,Hz.}
    \label{fig:contrast_vs_iterations_old}
\end{figure*}

\begin{figure}
    \centering
    \includegraphics[width=\linewidth]{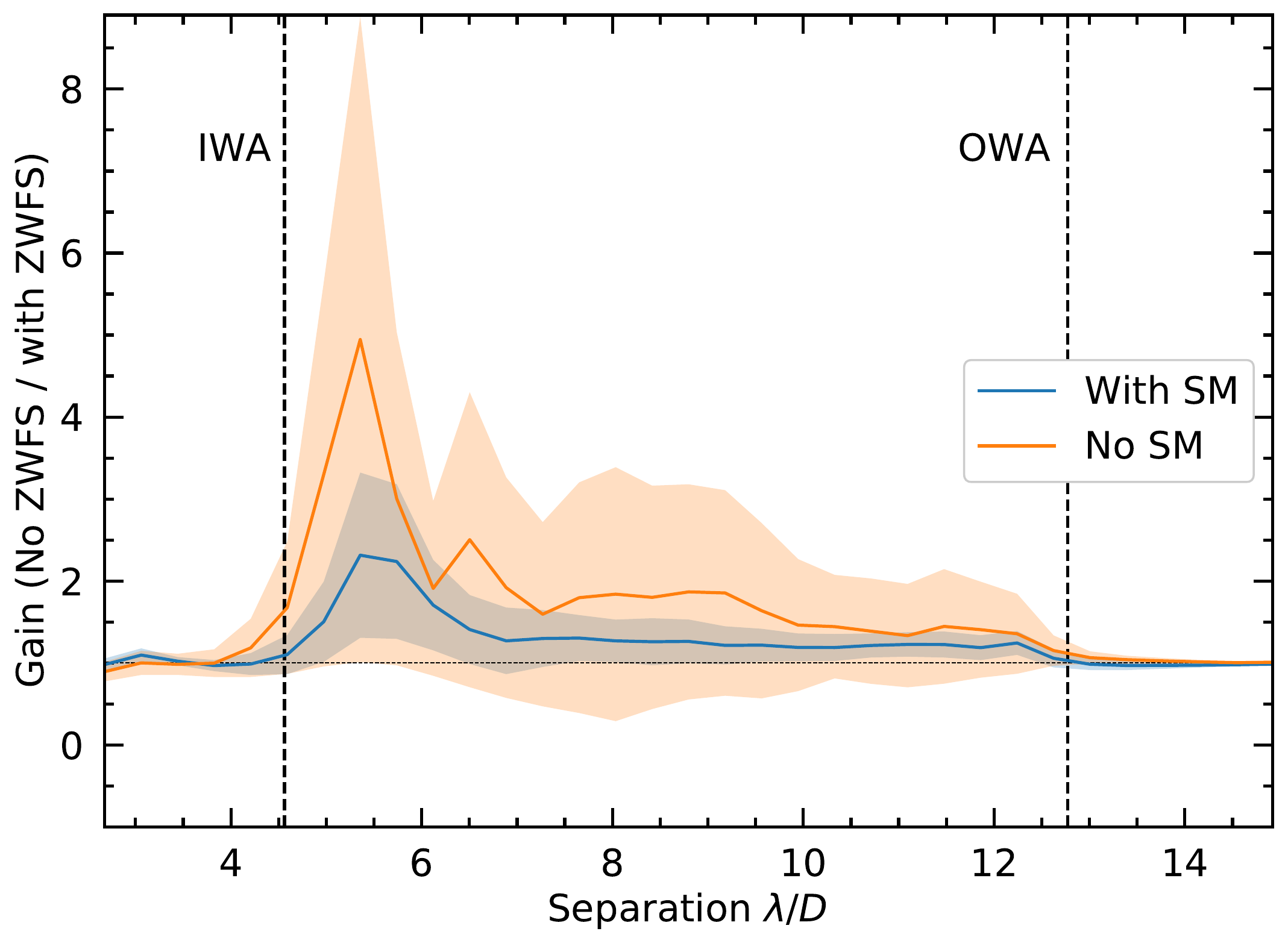}
    \caption{Contrast gain as a function of angular separation in the DH when the HOWFC loop is slowed down to 0.05\,Hz on purpose. Solid lines represent the ratio of the azimuthally averaged contrast between the cases without and with LOWFC. The filling colors show the standard deviation of the measurement. The dashed horizontal line is drawn at a gain of 1. The blue curve represents the gain between the values of the intervals (0) and (2) over the values of the intervals (1) and (3). The yellow curve corresponds to intervals (4) values.}
    \label{fig:gain_vs_sep_old}
\end{figure}

The contrast curve is presented in Fig.~\ref{fig:contrast_vs_iterations_old}. In this configuration, the LOWFC loop proves very efficient. While in interval (0) the contrast reaches a floor at $1.9\times10^{-7}$, turning on the LOWFC loop in (1) allows for an almost immediate contrast improvement by a factor of 1.5. This behavior is similar with the intervals (2) and (3). During interval (4), when the HOWFC loop is turned off, the drift of the uncontrolled DH is immediate, while the LOWFC loop manages to slow it down, limited by the reduced number of controllable modes. Overall, the contrast standard deviation over the intervals proves also to be greatly improved, from around $1.5\times10^{-7}$ to $1.5\times10^{-8}$ without and with LOWFC. The azimuthally averaged gain is represented in Fig.~\ref{fig:gain_vs_sep_old}, showing a gain of 2.5 and 6 with and without SM at short separations as in the previous experiments. But it also shows a gain larger than 1.5 at all separations, whether SM is on or off.

Under these different experimental conditions, we also show a successful concurrent operation. Similarly to the results in Sec.~\ref{sec:fast_results}, most of the improvement by the LOWFC is located at the inner radius of the DH, improving the contrast at short separations. By comparing the two setups, it is possible to extract properties of the control loops. In both setups, when the control loops are off, the contrast drift is immediate, emphasizing the presence of the perturbation which is partially corrected when the LOWFC loop is on.

The remaining contrast drift is due to the higher order perturbation. This perturbation is a main driver of the averaged contrast in the DH, and is out of reach for the LOWFS. This is particularly visible when looking at the intervals (0) in both setups, where the contrast levels only depend on the HOWFC speed. In similar conditions, the faster setup reaches a contrast level deeper than the slower setup by almost an order of magnitude, showing the impact of loop speed with the air turbulence on HiCAT.

With a faster speed than the HOWFC loop, the LOWFC loop can therefore correct for the low-order turbulence that evolves too fast for the high-order controller. As the leftover is more important in the slower setup, the contrast gain due to the ZWFS is greater and more obvious. On the contrary, with the faster setup, since the HOWFC is more efficient at correcting low-order aberrations, the gains are more moderate. Overall, there is a gain in using the LOWFS in both setups, showing that the loop combination does not generate cross-talks at the contrast levels in our experiments.

\section{Conclusions}
Wavefront error stability is a key parameter in the success of future space missions with exoplanet imaging capabilities. To improve the stellar light rejection by the coronagraph, WFSC appears essential to alleviate the requirements on the observatory stability for high-contrast observations. In particular, WFSC has been proven efficient to correct for different kinds of aberrations such as low-order drifts, or improve the image plane DH by optimizing the DM shapes. However, this was only done sequentially so far. Following the demonstration of DH stabilization with a ZWFS-based LOWFC loop in the rejected light of a Lyot coronagraph in \citetalias{Pourcelot2022}, we have here studied its concurrent operation with a HOWFC loop using different WFSs but the same DM. 

Using the HiCAT testbed for our experiments, we first study the impact of PW probes and SM commands on the LOWFC loop by projecting these commands on its controlled modes. We find a limited response, and considering the gain of the LOWFC loop of 0.01 as well as the different loop frequencies on HiCAT, the loop response to a single command is unlikely to modify the original command by more than 0.1\%. We find similar results for two kinds of probes for PW, whether single actuators poking or HiCAT probes. Then we run the HOWFC and LOWFC loops to dig or stabilize a focal-plane DH under the conditions of increased perturbation on the testbed. While the first loop is on, we maintain the contrast levels around $5\times10^{-8}$, showing no cross-talk between the two control loops. Even if the main DH contrast driver is due to high-order modes that evolve too fast for the HOWFC loop, it is possible to observe a contrast improvement with the LOWFC loop, with a factor of up to 1.5, and a factor of 5 when the HOWFC loop is on and off. To prove the efficiency of the LOWFC loop, we compare the performances with a reduced speed of the HOWFC loop. With the same IWA of $4.6\,\lsd$ this greatly degrades the achievable contrast, but emphasizes the capability of both loops to work together. A classical Lyot coronagraph with a smaller focal-plane mask and therefore a smaller IWA will be more sensitive to low-order aberrations, leading to an increased need for the LOWFC loop. Further investigations are required to find the best functioning point between the HOWFC and LOWFC loops for this setup. Preliminary tests have already been successfully performed on HiCAT with a knife-edge coronagraph, showing promising contrast stability results at shorter separations which will be detailed in a forthcoming paper \citep{Por2022}.

We validated our approach in a specific environment, in which the PW sensing runs at temporal frequencies very close to the LOWFC loop at 80\,Hz with a loop gain of 0.01 and a SM iteration every 2.5\,s. On average, a single LOWFC loop correction is applied at each PW probe. However, when we reduce the HOWFC loop speed, we manage to achieve a contrast of $10^{-7}$ with a very simple loop synchronization to avoid cross-talk. The asynchronous operations tend to converge toward DH solutions that are dependent on the LOWFC loop contribution, while we would prefer the two control loops to be as independent as possible. We currently operate HiCAT around contrasts of $5\times10^{-8}$. These loop interactions could create a bias in the wavefront estimation, thus representing a possible limitation when pushing the contrast to lower values. Regarding these points, it might be worth re-evaluating the asynchronous operations and considering an improved communication. This will enable adapting the LOWFC loop reference depending on the commands sent by the HOWFC loop, in a similar fashion as \citet{Guyon2020b}. In the scope of combining more loops, the asynchronous approach will remain applicable as long as there is no cross-talk between the loops. Further investigations on the loop communication will otherwise be required as the number of communications for offloading grows quadratically with the number of loops. 

We conducted our experiments in idealized conditions. First the tests are done in monochromatic light. While the ZWFS is expected to work relatively well with broadband light \citep{N'Diaye2013}, the DM DH commands will be different from those in monochromatic light. In further studies, we will investigate the stability of the parallel loop operations in broadband light, pending the delivery of a new broadband source. Second, our work is made with enough photons to correct as much turbulence as possible, the main limitation being the camera maximum frame rate. Assuming a JWST-like primary mirror, the equivalent magnitude of the monochromatic source on HiCAT would be about -6 in the visible. This is clearly an ideal case and the impact of signal-to-noise ratio on ZWFS measurements remains to be explored. We will investigate these aspects further, following \citet{Sahoo2022} who recently developed a novel approach to determine the optimal wavefront sensor exposure time, for a given contrast requirement at a given stellar magnitude.

A flux limitation will also likely increase the relevance of the LOWFS. In a photon-limited regime, a reduced photon flux requires longer exposure times on the science camera to keep the signal-to-noise ratio constant, leading to a possible reduction of the HOWFC efficiency. Since the ZWFS uses the light from the core of the source image, it collects more photons than the science camera. Its associated loop will therefore be able to run faster and correct for the aberrations beyond the HOWFC temporal bandpass. As emphasized by the configuration in Sec.~\ref{sec:slow_loops}, the low-order controller can then be complementary with the high-order controller.

Finally, these results have been obtained by degrading the HiCAT environment to generate larger drifts in an uncontrolled DH. In this context, we do not reach the optimal performance of HiCAT that is way more stable in nominal conditions. In particular, with the injected turbulence, here we are limited in contrast by the speed of the HOWFC loop that runs at 0.4\,Hz. Using LOWFC, the contrast improvement is limited to separations close to DH IWA, where the control regions of both loops is smoothly overlapping. It would be relevant to study how far out in the focal plane the LOWFC loop could effectively control by considering more than the current 20 Zernike modes. This could allow a gain in the temporal bandpass of correction as the LOWFC loop runs orders of magnitude faster than the HOWFC loop and alleviates the latter. 

At the moment, our demonstration of the concurrent use of LOWFC and HOWFC loops at the 5$\times$ 10\up{-8} contrast point represent a first milestone towards concurrent operations with a 10\up{-10} contrast goal. Several aspects in our current experiment such as the selected control modes, the accuracy of the DM behavior modeling, or the sensor sensitivity, will be further explored in the regime of wavefront error fluctuations down to the picometric level to further advance exo-Earth imaging with concurrent loops.

This combination of several control loops is a necessary step toward a system-level demonstration of a future high-contrast instrument for exoplanet imaging with a future large space observatory. To stabilize the whole range of aberrations that can disturb the observations, these control loops have to be associated to others. These include a loop dedicated to tip/tilt correction, with a dedicated steering mirror already implemented on HiCAT but currently unused, another to primary mirror segment alignment or one dedicated to vibration rejections. Fully operating and understanding this setup will help develop future instrumentation for exoplanet and more particularly exo-Earth imaging.

\begin{acknowledgements}
R.P. acknowledges PhD scholarship funding from R\'egion Provence-Alpes-C\^ote d'Azur and Thales Alenia Space.
The authors are especially thankful to the extended HiCAT team (over 50 people) who have worked over the past several years to develop this testbed. This work was supported in part by the National Aeronautics and Space Administration under Grant 80NSSC19K0120 issued through the Strategic Astrophysics Technology / Technology Demonstration for Exoplanet Missions Program (SAT-TDEM; PI: R. Soummer).  E.H.P. was supported by the NASA  Hubble Fellowship grant HST-HF2-51467.001-A awarded by the Space Telescope Science Institute, which is operated by the Association of Universities for Research in Astronomy, Incorporated, under NASA contract NAS5-26555. I.L. acknowledges the support by a postdoctoral grant issued by the Centre National d'Études Spatiales (CNES) in France.

\end{acknowledgements}

\bibliographystyle{aa}
\bibliography{biblio}

\end{document}